# THERMAL RADIO EMISSION:
# THE BRIGHTNESS TEMPERATURE AND THE SPECTRAL INDEX OF RADIO EMISSION

Fedor V. Prigara

*Institute of Microelectronics and Informatics, Russian Academy of Sciences,*
*21 Universitetskaya, 150007 Yaroslavl, Russia*

The condition of radio emission is proposed, on the base of which the theory of themal radio emission for gaseous disk is developed. This theory explains the radio emission spectra of known types of extended radio sources, located beyond the Solar planetary system. Besides, the thermal radio emission spectra of Venus and Jupiter are explained.

In this paper the thermal radio emission of non-uniform gas is considered. For instance, the gas may be placed in a field of gravity. We suggest, that the thermal radio emission with a wavelength $\lambda$ is emitted by the gaseous layer, for which the next condition of radio emission is valid:

$$\lambda = l = \frac{1}{n\sigma}. \tag{1}$$

Here $l$ is the mean free path of emitting particles, $n$ is the concentration of emitting particles (atoms, ions or molecules), $\sigma$ is the effective scattering cross-section.

It is supposed also, that the thermal radiation is emitted only in the direction, in which the concentration of emitting particles decreases, but not in the opposite direction. The intensity of thermal radio emission is given by Rayleigh-Jeans formula

$$I_\nu = 2kT\frac{\nu^2}{c^2}, \tag{2}$$

where $T$ is the temperature of emitting gas, $k$ is the Boltzmann constant, $c$ is the velocity of light and $\nu=c/\lambda$ is the frequency of radio emission [1].

It is not supposed, howevere, that an emitting gaseous layer absorbes all the incident radiation with the same frequency as that of emitted radiation. On the contrary, an emitting gaseous layer may be quite transparent for the emission with frequency $\nu$. We notice, in this context, that the Kirchhoff's consideration [2], which establishes the relation between the emission and absorption coefficients, is not valid for non-uniform systems. In the case of non-uniform system, let us say a planet with its atmosphere, the radiation with given frequency $\nu$ may be absorbed by one part of the system (for instance, by the surface of planet) and emitted by the other part (for instance, by the definite gaseous layer in the atmosphere of planet). The energy transfer from one part of system to the other is realized by the means of convection, thermal conductivity or emission in the other range of frequencies. In this case Kirchhoff's law is valid for the non-uniform system in whole (a planet with its atmosphere), but not for its parts.

In order to elucidate the condition of emission (1) let us consider the Einstein's coefficients $A_{21}$, $B_{21}$ and $B_{12}$, where $A_{21}$ and $B_{21}\rho$ are the numbers of transitions from the energy level 2 to the energy level 1 per unit time, corresponding to a spontaneous and an induced emission respectively, $\rho$ is the energy density of blackbody radiation, and $B_{12}\rho$ is the number of transitions from the energy level 1 to the energy level 2 per unit time, the energy of level 2 being higher than the energy of level 1 [2,3].

In the case of non-degenerated levels the Einstein's coefficients obey the relation

$$\exp(\hbar\omega/kT) = (A_{21} + B_{21}\rho)/(B_{12}\rho), \qquad (3)$$

where $\hbar$ is Planck's constant, $\omega$ is the circular frequency of light, $T$ is the temperature, and $k$ is Boltzmann's constant. Taking into account the relation $B_{12}=B_{21}$, we obtain from the equation (3)

$$A_{21}/(B_{21}\rho) = \exp(\hbar\omega/kT) - 1. \qquad (4)$$

If $\hbar\omega \ll kT$ which is the case for thermal radio emission, then $A_{21} \ll B_{21}\rho$. Therefore the contribution of a spontaneous emission to thermal radiation in radio wavelength range may be neglected. Thus thermal radio emission is produced by induced emission, e.i. the emission of maser type.

Retaining to the equation (1) we can interpret the mean free path $l$ as the size of a molecular resonator. The emitting gaseous layer contains many molecular emitters of size $l$. Each molecular emitter produces the coherent radiation , thermal radiation being incoherent sum of radiation produced by individual emitters .

Let us apply the above formulated condition of emission to the atmosphere of planet. In this case the concentration of molecules $n$ continuosly decreases with the increase of height $h$ in accordance with barometrical formula

$$n = n_0 \exp(-\frac{mgh}{kT}), \qquad (5)$$

where $m$ is the mass of molecule, $g$ is the gravity acceleration, $k$ is the Boltzmann constant [4]. Here the temperature of atmosphere $T$ is supposed to be constant. In fact the temperature changes with the increase of height, so the formula (5) describes the changement of molecule concentration in the limits of layer, for which we can consider the temperature to be approxymately constant ( the temperature changes with the increase of height essentially slower than the concentration of molecules ).

If the emitting particles are the molecules of definite sort, then their concentration is monotonously decreasing with the increase of height. Therefore, according to the condition of emission (1), the radiowaves with wavelength $\lambda$ are emitted by gaseous layer, located at well defined height $h$ in atmosphere. Thus, the relation between the brightness temperature of radio emission and wavelength $T(\lambda)$ reproduces ( partially or in whole ) the temperature section across the atmosphere $T(h)$.

This is the case for the Venus atmosphere. Here the emitting particles are the molecules of $CO_2$. This oxide of carbon is the main component of Venus atmosphere, its contents being 97%. The data concerning the brightness temperature of radio emission from Venus are summerized in [5].

Assuming $s \approx 10^{-15}$ cm$^2$ and using the condition of emission (1), we can find the concentration of emitting molecules $n$ in the gaseous layer, which emittes the radiowaves with given wavelength $\lambda$. Then, with the help of formula (5) and the data upon the pressure and the temperature in the lower layer of Venus atmosphere [4], we can establishe the height of emitting layer in atmosphere. This procedure gives the temperature section of Venus atmosphere showen in figure 1. It is not in contradiction with the data received by the means of spacecrafts Venera, since on these apparatus the temperature was measured only in the limits of troposphere, up to the height 55 km.

We observe, that the temperature structure of Venus atmosphere is similar to those of Earth's atmosphere [4]. Because of this analogy to denote the various layers of Venus atmosphere the same terms were used here as those which are usually applied for the Earth's atmosphere. Up to date there was no acceptable explanation for the decrease of brightness temperature in the decimeter range, corresponding in our consideration to the emission of mesopause region.

Quite similarly the brightness temperature of Jovian thermal radio emission in the range 0.1-4 cm as a function of wavelength reproduces the temperature section of Jovian atmosphere. See [6], where the data upon the microwave emission of Jupiter are represented, and also [1,5]. In this case the temperature strucrure of atmosphere also is similar to those of Earth's atmosphere : there are two minimums of temperature ( tropopause and mesopause ) and one intermediate maximum (mesopeak ).

The observational data concerning the thermal radio emission of Mars, Saturn, Uranus and Neptune [1] are not so complet as in the case of Venus and Jupiter. These data, however, are in agreement with Earth type temperature structure of atmosphere. In the case of Mars the measurements of pressure and temperature in the lower layer of atmosphere are available [5], so one can reproduce using the radio emission data the temperature section of Mars atmosphere. Here the emissing particles are the $CO_2$ molecules as in the case of Venus atmosphere.

The intermediate maximum of temperature in the region of mesopeak perhaps can be explained by absorbtion and next reemission of infrared radiation transfered from the lower layers of atmosphere. It is worthwhile to notice, that the temperature in the region of mesopeak is usually close to the temperature of planet surface.

Consider now a gaseous disk with thickness $d$ and radius $R$, in which the regular radial convection is realized : in the medial plane of disk gas flows to the center of disk and near the upper and lower surfaces of disk returns to periphery. The velocities of convection flows lie in planes, containing the normal to disk plane.

Since the total number of particles is conserved, the bulk concentration of particles (ions) $n$ decreases with the increase of radius $r$ as

$$n \propto r^{-1} . \qquad (6)$$

At the upper and lower surfaces of disk and also at $r=R$ the concentration of particles (ions) decreases to some small value.

Consider now the thermal radio emission of such disk in the direction of normal to disk plane. According to the emission condition (1), radiowaves with frequency $\nu$ are emitted by the region of disk with $r \leq r_n$, where $r_n$ is the value of radius, for which the bulk concentration of ions $n$ satisfy the relation (1). The spectral density of radio emission flux at frequency $\nu$ is given by formula

$$F_n = pI_n j_n^2 = pI_n \frac{r_n^2}{a^2}, \qquad (7)$$

where the intensity $I_n$ is given by Rayleigh-Jeans formula (2) and $a$ is the distance from gaseous disk (nebula) to the detector of radio emission.

From relations (1) and (6) we obtain

$$r_n \propto l \propto n^{-1}. \qquad (8)$$

Assuming the temperature $T$ to be constant in the whole volume of gaseous disk, from relation (7) we find that if $r_n < R$, then the flux $F_\nu$ is independant of frequency: $F_\nu$=const. And if $r_n = R$ (that corresponds to the longwave range $\nu<\nu_0$), then the flux $F_\nu$ is proportional to the square of frequency:

$$F_n \propto I_n \propto n^2. \qquad (9)$$

Exactly such spectra are observed for the most of planetary nebulae [7]. For instance, in the case of planetary nebula NGC 6543 the boundary wavelength deviding the regions with spectra $F_n \propto n^2$ and $F_n = const$, equales $l_0$=30 cm, for NGC 7027 $l_0$=10 cm and for IC 418 $l_0$=20 cm.

The spectrum $F_n \propto n$ observed for the series of compact nebulae, e.g. Vy 2-2, can be explained by the decrease of temperature with the increase of radius in accordance to the law

$$T \propto r^{-1}. \qquad (10)$$

Planetary nebulae are as a rule stationary, not expanding objects. A motion of details, which was interpreted as the consequence of expansion of nebula, one can attribute to the radial convection. The interpretation of spectral lines splitting as the consequence of expansion leads to some difficulties [7]. It is worthwhile to remark that the existing theory of Doppler effect is unable to explain the non-uniform shift of spectral lines which is often observed, e.g. in the spectra of novae after the flare. See [8], where the spectrum of Hercules Nova 1934 after the flare is described with the stationary line $H_a$ and periodically shifting lines $H_e$ and $H_v$ in the violet range of spectrum.

The recente observations show the jet structure of some planetary nebulae [9]. The jets can be interpreted as a result of interaction of convective flows with two-dimentional magnetic field of planetary nebula. The last forms and stabilizes a gasious disk discussed above. The radio image of planetary nebula Hb 12 obtained at the wavelength $\lambda$=6 cm (see [7],Ch.1) shows the extented detailes following the force lines of magnetic field for two -dimentional dipole.

It follows from equation (8) that the angular size of radio source depends on the wavelength of radio emission, namely the angular dimensions of radio source are proportional to the wavelength $\lambda$. The wavelength dependence of radio source size is confirmed by observations [10,11]. For instance, in the case of galaxy M31 the radius of the central radio core is 3.5 arcmin at the frequency 408 MHz ($\lambda$=74cm) and less than 1.3 arcmin at the frequency 1407 MHz ($\lambda$=21 cm) [11]. The $\lambda^2$ dependence of source size observed in the case of Sgr A* [10] can be obtained if the effect of magnetic field on the flow of ionized gas is taken into account. At small values of radius this effect changes the equation (6) for equation

$$n \propto r^{-1/2} \qquad (11)$$

as it follows from geometrical considerations. The radio source Sgr A* has a flat spectrum, corresponding to the relation (10), since in this case $r_n \propto n^{-2}$.

In the case of supernovae remnants we also can apply the theory of thermal radio emission for gaseous disk. The main difference is that instead of the condition of temperature constancy here often the condition of gas pressure constancy is valid

$$P = nkT = const, \qquad (12)$$

where *n* as above is the number of particles (ions) per unite volume, *P* is the pressure and *T* is the temperature of gas, and *k* is the Boltzmann's constant. The condition *P=const* and relation (4) lead to the dependance of gas temperature upon radius *r* :

$$T \propto r. \qquad (13)$$

In the consequence of this relation the flux of radio emission depends on frequency ν as follows

$$F_n \propto T n^2 r_n^2 \propto n^{-1}. \qquad (14)$$

The observed spectra of radio emission for supernovae remnants can be interpreted as the combination of spectra $F_\nu = const$ and $F_\nu \propto \nu^{-1}$. For instance, in the case of radio source Cassiopeia A in the wavelength interval 3 cm<λ<25 cm the law (14) is valid and in the interval 1.2 m <λ<10 m the spectrum is $F_\nu = const$ [12].

Here we must make a remark that Cassiopeia A as well as planetary nebulae is the stationary, non expanding object. This fact explains the absence of parent supernova flare which could generate this nebula. Such flare of supernova should take place in 17 century [12,13]. The connection between the optical details of this nebula and the radial convection is obvious : the details appear "from nothing" and disappear, sometimes decaying into smaller parts (see [12]). Described in [12] the change of radio flux from Cassiopeia A can be interpreted as periodical effect with the period *T*≥100 years. Besides, the estimations of expansion time scale made in various ranges ( optical, X-rays and radio ) are not consistent each with other [13].

Since the intensity of thermal radio emission is proportional to temperature, the increase of temperature in accordance to relation (13) will cause the enlargement of radio brightness on the edge of nebula. Such effect is observed for the most of supernovae remnants, in particular for Cassiopeia A [12]. However there are supernova remnants with the more uniform distribution of radio brightness and respectfully with more flattened spectra $F_\nu \approx const$ corresponding the conditon $T \approx const$ [12,14]. Notice, that the correlation between the spectral index and the radial distribution of radio brightness is in contradiction with the synchrotron theory of radio emission for supernova remnants.

The theory of thermal radio emission for gaseous disk describes also the radio emission of radio galaxies and quasars. As well as for supernova remnants here the spectra $F_\nu \propto \nu^{-1}$ and $F = const$ are observed. For instance, the radio spectrum of Virgo A is the combination of these two spectra, similarly to the spectrum of Cassiopeia A [1]. For the quasar NRAO 530 $T \gg const$, $F_\nu \gg const$, and $r_1 \mu 1$ in the range of wavelengths 0.3 cm to 6.3 cm [15].

The radio spectrum of radio galaxy Hercules A (radio source 3C 348) is described by law (12) in the wavelength interval λ=10-170 cm.

The spectrum $F_\nu \propto \nu^{-1}$ is typical for extented radio sources with jets. Compact radio sources with angular dimensions smaller than 0.1 arcmin as a rule demonstrate the spectrum $F_\nu \approx const$ [16]. The last spectrum is also typical for BL Lac objects. It is worthwhile to notice that theory leading to the law (14) is still valid in the case of sector of disk (instead of whole disk), so one can apply this theory to the jets.

Thus there are two main types of extended radio sources. I type radio sources are characterized by the stationary convection in gravitational or magnetic field with approximately uniform distribution of gas temperature. II type radio sources have the outflows of gas with approximately uniform distribution of gas pressure.

I type radio sources include planetary nebulae, supernova remnants with pulsars (center-brightened shells) [12,14] and compact sources corresponding to galactic nuclei. II type radio sources include the most of supernova remnants (edge-brightened shells) [12,14] and active galactic nuclei with jets.

It is possible that the radio source core belongs to the I type radio sources and the jets belong to the II type sources, as in the case of radio galaxy PKS1549-79 [16].

There are also compact radio sources with the distribution of temperature described by equation (10) , such as compact planetary nebulae of Vy 2-2 type and Sgr A* . They may be classified as III type radio sources .

Radio spectrum of galaxy M31 is described by the law (14) in the wavelength interval λ=20-170 cm.[11] .

Radio emission of diffuse nebulae is similar to that of planetary nebulae [1].

Thus we conclude that the radio spectra of known types of extended radio sources located beyond the Solar planetary system are described by the theory of thermal radio emission for gaseous disk which was developed above. The exceptions include maser sources and pulsars.

The possibility to apply the unified model of gaseous disk to gaseous nebulae and active galactic nuclei is justified by the similarity of optical emission spectra . Besides , there is a non-uniform shift of the spectral lines in the spectra of quasars and radio galaxies , similar to those in the spectra of planetary nebulae .

The thermal origin of emission in the case of active galactic nuclei is supported by the recent observation of the [OIII] lines polarization in the spectrum of radio galaxy PKS1549-79 coincident with the polarization of optical continuum [16] .

----------------------------------------------------------

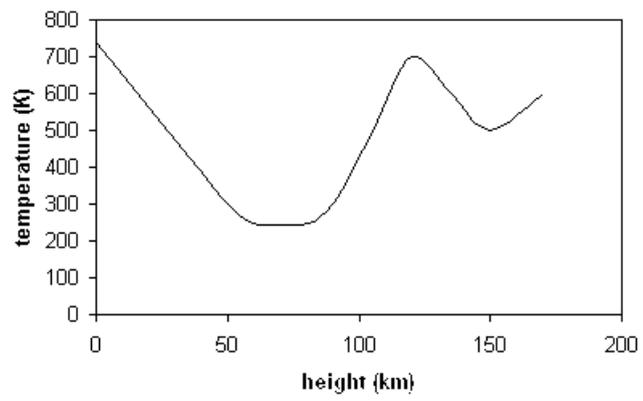

FIG.1. Temperature as a function of height in Venus atmosphere.